\documentclass{article}[fullpage]
\usepackage{amsfonts,amsmath,amsthm,braket,framed}
\usepackage[backref,hidelinks]{hyperref}
\usepackage[margin=2cm]{geometry}
\usepackage{authblk}
\usepackage{algorithm}
\usepackage{algpseudocode}
\usepackage{multirow}

\newcommand{\C}{\mathbb{C}}
\newcommand{\I}{\mathbb{I}}

\newcommand{\Z}{\mathbb{Z}}

\newcommand{\cL}{\mathcal{L}}
\newcommand{\cS}{\mathcal{S}}

\newcommand{\ks}{\ket{s}}

\newtheorem{definition}{Definition}

\newtheorem{thm}{Theorem}
\newtheorem{lemma}{Lemma}

\title{Bases for optimising stabiliser decompositions of quantum states}

\author[1]{Nadish de Silva\footnote{Corresponding author: \url{ndesilva@sfu.ca}.}}

\author[1]{Ming Yin}
\author[2]{Sergii Strelchuk}
\affil[1]{Department of Mathematics, Simon Fraser University}
\affil[2]{Department of Applied Mathematics and Theoretical Physics, University of Cambridge}

\begin{document}

\date{} 
\maketitle

\begin{abstract}
Stabiliser states play a central role in the theory of quantum computation.  For example, they are used to encode computational basis states in the most common quantum error correction schemes.  Arbitrary quantum states admit many \emph{stabiliser decompositions}: ways of being expressed as a superposition of stabiliser states.  Understanding the structure of stabiliser decompositions has significant applications in verifying and simulating near-term quantum computers.

We introduce and study the vector space of linear dependencies of $n$-qubit stabiliser states.  These spaces have canonical bases containing vectors whose size grows exponentially in $n$.  We construct elegant bases of linear dependencies of constant size three.  

Critically, our sparse bases can be computed without first compiling a dictionary of all  $n$-qubit stabiliser states.  We utilise them to explicitly compute the stabiliser extent of states of more qubits than is feasible with existing techniques.  Finally, we delineate future applications to improving theoretical bounds on the stabiliser rank of magic states.
\end{abstract}

\section{Introduction and background}

\paragraph{Stabiliser states}  \emph{Stabiliser states} \cite{gottthesis, gottesman1999heisenberg} are a key class of states that include many of the important states in quantum computing, e.g.\ the Bell state and the GHZ state.  They play a central role in the theory of quantum error correction as they are used to encode quantum data over physical qubits.  Deeper understanding of the structure of stabiliser states \cite{dehaene2003clifford,garcia2014geometry,hu2022improved} will have applications across quantum information.  

In this work, we ask the foundational question: what are all of the linear dependencies of stabiliser states?  Given the essential role of superpositions of stabiliser states in quantum information and the mathematically fundamental nature of the question posed, the answer will find applications across quantum information.  We are particularly motivated by applications to the development of classical algorithms for the simulation of quantum circuit families.

\paragraph{Classical simulation algorithms} The study of classical algorithms for the simulation of quantum circuit classes is a promising approach to one of the central problems of quantum computation: identifying computational settings wherein quantum computers offer an advantage.  Quantum circuit classes which admit efficient classical simulation offer no advantage.  Determining the minimal resources needed to boost the computational power yields insight into the nature of quantum advantage.  At the same time, classical simulation algorithms are of immediate practical importance as they are required to verify experimental prototypes of quantum computers.

The Gottesman-Knill algorithm \cite{gottesman1999heisenberg,aaronson2004improved} is the foremost example of a classical simulation algorithm.  It allows a classical computer to efficiently emulate the class of Clifford circuits.  Magic states \cite{shor1996fault,gottesman1999}, e.g.\ the $\ket T$ state \cite{bravyi2005universal}, were identified as minimal extra resources that promote Clifford circuits to full quantum universality.

\paragraph{Stabiliser rank}  The Clifford+$T$ circuit model forms the basis of a classical simulation scheme capable of emulating universal quantum computation: the \textit{stabiliser rank simulation} of Bravyi-Smith-Smolin \cite{bravyi2016trading}.  It is the most widely studied algorithm for classically simulating quantum circuits dominated by Clifford gates and, in the case of more general Clifford+$T$ circuits, performs competitively compared to tensor-network-based 
 methods~\cite{markov2008simulating}. 

In this scheme, an arbitrary quantum circuit is first expressed as a stabiliser circuit of Clifford gates acting on its input, together with quantum resources in the form of $n$ nonstabiliser magic states $\ket{T}^{\otimes \, n}$, where $\ket{T} = T\ket{+}$.  The magic states are expressed as a superposition of stabiliser states: $\ket{T}^{\otimes \, n} = S_n \, \vec{x}$ where $S_n$ is a matrix whose columns are the $n$-qubit stabiliser vectors $\ket{s_i}$.  Such a vector $\vec{x}$ is a \textit{stabiliser decomposition} of $\ket{T}^{\otimes \, n}$ and, crucially, it is not unique.  Indeed, any state has infinitely many possible stabiliser decompositions.

This algorithm can approximate the output of a quantum circuit with a runtime proportional to the number $\chi$ of nonzero $x_i$ for the stabiliser decomposition used.  The minimal $\chi$ for which a state can be expressed as a superposition of $\chi$ stabiliser states is the \emph{stabiliser rank} of the state.     

Finding short stabiliser decompositions of magic states is a very difficult and urgent open research question \cite{qassim2021classical,qassim2021improved}.  At present, the only method for finding low-rank stabiliser decompositions is the one originally provided in \cite{bravyi2016trading}.  It is an algorithm based on a random walk and is limited by efficiency constraints to a maximum of six-qubit states.  This is due to the vast number of combinations of stabiliser states possible and the lack of any structured approach.  As a consequence, the effectiveness of the stabiliser rank to simulate medium-term devices is hampered.

The stabiliser rank of an important non-magic family of states, Dicke states~\cite{dicke1954coherence}, has also been investigated~\cite{vinkhuijzen2021limdd}.  These states have a number of applications in simulating quantum networking algorithms, game theory, QAOA and quantum  metrology~\cite{ prevedel2009experimental, ozdemir2007necessary, toth2012multipartite, childs2000finding, farhi2014quantum, hadfield2019quantum}. Currently, there is an effort to design low-depth quantum circuits that prepare these states efficiently on a quantum computer~\cite{bartschi2019deterministic, bartschi2022short}.  

It is also notoriously difficult to prove that the stabiliser rank of $\ket{T}^{\otimes n}$ grows exponentially in $n$.  It must grow at least superpolynomially unless $\mathsf{P}=\mathsf{NP}$ \cite{huang2020explicit}.
Despite the sophisticated areas of mathematics brought to bear on this problem --- including ultra-metric matrices \cite{bravyi2019simulation}, higher-order Fourier theory \cite{labib2022stabilizer}, number theory/algebraic geometry  \cite{lovitz2022new}, probability theory \cite{mehraban2023lower}, and Boolean analysis \cite{peleg2022lower} --- only a linear lower bound has been achieved.



The obstructions to progress on these, and other \cite{howard2017application,heinrich2019robustness, zurel2020hidden}, questions may be attributable to an underdeveloped understanding of the geometric structure of the set of stabiliser states.

\paragraph{Stabiliser extent} 
Computing the stabiliser rank (and its approximation \cite{bravyi2016improved}) has proved to be a notoriously difficult task and this necessitates the search for upper- and lower-bounds which are less computationally costly. This motivated the authors of~\cite{bravyi2019simulation} to introduce the \emph{stabiliser extent}: a more computationally feasible quantity that upper bounds the approximate stabiliser rank \cite[Theorem 1]{bravyi2019simulation}. The stabiliser extent, denoted as $\xi$ and evaluated on a state $|\psi\rangle$ is defined as:
$$
    \xi (|\psi\rangle) = \min{\lVert \vec x\,\rVert ^2_1}\ \text{ s.t. }|\psi\rangle = S_n \vec x.
$$
This quantity can be efficiently minimised using a second order cone program.  For the case where $\ket{\psi} = \ket{\psi_0}^{\otimes k}$ is a tensor product of $k$ copies of a single 1-, 2-, or 3-qubit state $\ket{\psi_0}$, then $\xi(\ket{\psi}) = \xi(\ket{\psi_0})^k$ ~\cite[Proposition 1]{bravyi2019simulation}.  This renders the calculation of $\xi(\ket{\psi})$ even more efficient by reducing it to, at most, the $n=3$ case.  However, the stabiliser extent was shown to be nonmultiplicative in general~\cite{heimendahl2021stabilizer}.  As $n$ grows, the calculation of $\xi$ becomes much more difficult since the size of $S_n$ grows exponentially.  We demonstrate below how our techniques can be used to compute the stabiliser extent in the regime of large states.

The limits of classical simulability for Clifford+$T$ computations using sampling-based estimators of the stabiliser extent have been studied~\cite{kocia2022more}. One can also further extend the above notions to mixed states via the convex roof extensions~\cite{seddon2021quantifying}. The knowledge of the extent for general states in these cases can be translated into an efficient classical algorithm that can approximately sample from the probability
distribution which is generated by a sequence of Pauli measurements~\cite{seddon2021quantifying}.

\paragraph{Main results}  Our contribution below is to achieve the deeper understanding of the mathematical structure of stabiliser decompositions required to improve simulation efficiency and provide the technical tools for approaching hitherto intractable theoretical problems.
\begin{enumerate}
    \item We define a vector space $\cL_n$ of formal linear combinations of stabiliser states with vanishing sum.  The elements of $\cL_n$ are in correspondence with the stabiliser decompositions of any fixed $n$-qubit state (Section 3).
    \item We construct particularly nice bases of these spaces using only linearly dependent triples (Section 4).
    \item We show how to explicitly compute these bases for use in concrete applications (Section 5).
    \item We demonstrate an application of our methods to the computation of the stabiliser extent of large states, i.e.\ beyond the regime of 5-qubit states to which the state-of-the-art methods are limited (Section 6 and Table \ref{stabextenttable}).
\end{enumerate}

The code and data that support Sections 5 and 6 are publicly available at \url{https://github.com/ndesilva/stabiliser-decomp-bases}.


\section{Preliminaries}

\paragraph{The stabiliser formalism}  The $n$-qubit stabiliser formalism is based on the Pauli group, which is generated by the matrices $Z, X$ with the action on computational basis states: \begin{align*}
Z^{\vec{p}} \ket{\vec{z}\,} &= (-1)^{\vec{z} \cdot \vec{p}}\ket{\vec{z}\,}, \\
X^{\vec{q}} \ket{\vec{z}\,} &= \ket{\vec{z} + \vec{q}},\end{align*} for $\vec{z},\vec{p},\vec{q} \in \Z^n_2$ where $Z^{\vec{p}}$ denotes $Z^{p_1} \otimes \ldots \otimes Z^{p_n}$.  Given a string of $2n$ bits $(\vec{p},\vec{q}) \in \Z_2^{2n}$, the corresponding Pauli is: $$W(\vec{p},\vec{q}) = (-i)^{\vec{p} \cdot \vec{q} \text{ mod 4}} Z^{\vec{p}} X^{\vec{q}}.$$  We consider $\Z_2^{2n}$ a symplectic vector space over $\Z_2$ with the symplectic product $$[(\vec{p}_1,\vec{q}_1),(\vec{p}_2,\vec{q}_2)] = \vec{p}_1 \cdot \vec{q}_2 - \vec{p}_2 \cdot \vec{q}_1.$$  Then $W(\vec{p}_1,\vec{q}_1)$ and $W(\vec{p}_2,\vec{q}_2)$ commute if and only if $[(\vec{p}_1,\vec{q}_1),(\vec{p}_2,\vec{q}_2)] = 0$.

\paragraph{Lagrangian subspaces}  A Lagrangian subspace is a vector subspace of $\Z_2^{2n}$ that is maximal among those with the property that any two members have vanishing symplectic product.  They are typically specified by a set of $n$ vectors $\{(\vec{p}_1,\vec{q}_1),\ldots,(\vec{p}_n,\vec{q}_n)\}$ which, besides having pairwise vanishing symplectic product, is linearly independent.  

\paragraph{Stabiliser states}  Having fixed a Lagrangian subspace, a \emph{stabiliser state} $\ks$ is specified by a further $n$ bits $(\lambda_1,\ldots,\lambda_n) \in \Z_2^n$ via the eigenvalue equations $(-1)^{\lambda_j} W(\vec{p}_j,\vec{q}_j)\ks =  \ks$ for $j = 1, \ldots, n$.   The explicit amplitudes of stabiliser states were first determined by Dehaene and De Moor \cite{dehaene2003clifford}:  there is a stabiliser state $$\ket{A, q, l} = \frac{1}{|A|} \sum_{\vec{z\,}\in A} (-1)^{q(\vec{z\,})} i^{l(\vec{z\,})} \ket{\vec{z\,}}$$ for every affine subspace $A \subset \Z_2^n$, quadratic form $q: \Z_2^n \to \Z_2$, and linear function $l: \Z_2^n \to \Z_4$.

Note that these equations specify a state $\ks$ only up to phase.  We therefore select one member of each phase equivalence class by defining $\cS$ to be the set of stabiliser states with normalised phase, e.g.\ by fixing the first (i.e. having smallest computational basis bitstring, in the order of binary integers) nonzero amplitude to be real and positive.  We denote by $\cS_N$ the set of \emph{noncomputational} stabiliser states: $\cS \setminus \{\ket{\vec{z}} : \vec{z} \in \Z^n_2\}$. By the \emph{support} of a stabiliser state $\ks$, we mean $\{\ket{\vec{z}\,} : \vec{z} \in \Z^n_2, \braket{\vec{z} \,|\, s} \neq 0\}$; by the \emph{size} of its support, we mean the number of elements of this set.

\paragraph{Linearly dependent triples}  Garcia et al.\ \cite{garcia2014geometry} first explored geometric aspects of the set of stabiliser states and studied the problem of classifying the linearly dependent triples of stabiliser states.  That is: those triples of stabiliser states $\ket {s_1}, \ket {s_2}, \ket {s_3}$ such that there exists $x_1,x_2,x_3 \in \mathbb{C}$, not all zero, such that $$x_1 \ket{s_1} + x_2 \ket {s_2} +  x_3 \ket {s_3} = \vec 0.$$  It was claimed that for nonparallel triples, up to relabelling the states, $|{\braket{s_1|s_2}}|=|{\braket{s_1|s_3}}|=2^{-1/2}$, $\braket{s_2|s_3}=0$.  This classification was found to be incomplete by Hu-Khesin \cite{hu2022improved} and, independently, by the present authors. 


\section{The space of linear dependencies}\label{lindeps}

 We formalise the idea of a linear combination of stabiliser states and of a linear dependency of stabiliser states.  These notions facilitate study of the space of stabiliser decompositions of any state.

\begin{definition}
A \emph{(formal) linear combination of stabiliser states} is a function $\alpha: \cS \to \C$.
\end{definition} \noindent Such a linear combination of stabiliser states naturally yields the vector $$\sum_{\ks \in \cS} \alpha(\ks) \ks \in \C^{2^n}.$$  

\begin{definition}
A \emph{stabiliser decomposition} of a state $\ket{\psi}$ is a linear combination $\alpha \in \C^\cS$ such that $$\ket{\psi} = \sum_{\ks \in \cS} \alpha(\ks) \ks.$$
\end{definition} \noindent The \emph{support} of a linear combination $\alpha$ is $\alpha^{-1}(\C \setminus \{0\})$.  

  Given a second stabiliser decomposition $\beta$ of $\ket{\psi}$, $\alpha$ and $\beta$ differ by a linear combination $\gamma$ with the property $\sum_{\ks \in \cS} \gamma(\ks) \ks = \vec{0}.$  

\begin{definition}
A \emph{linear dependency} of stabiliser states is a linear combination $\alpha \in \C^\cS$ such that $$\sum_{\ks \in \cS} \alpha(\ks) \ks = \vec 0.$$
\end{definition} \noindent

\begin{definition}
The set $\cL_n$ is the subset of $\C^\cS$ defined as the set of all linear dependencies of $n$-qubit stabiliser states. 
\end{definition} \noindent  It is easy to see that the $\cL_n$ form complex vector spaces.  By understanding the structure of these vector spaces, we acquire basic tools for working with superpositions of stabiliser states, e.g.\ stabiliser decompositions.  

\begin{lemma}The set of all stabiliser decompositions of a state $\ket{\psi}$ is the affine space $\tilde{\psi} + \cL_n$ where $\tilde{\psi}$ is any decomposition of $\ket{\psi}$, e.g.\ its expansion in the computational basis. \end{lemma}  

\begin{lemma} The dimension of $\cL_n$ is $|\cS| - 2^n = |\cS_N|$.\end{lemma}
\begin{proof}
The space $\cL_n$ is the kernel of the linear transformation $e: \C^\cS \to \C^{2^n}$ that evaluates a formal linear combination $\alpha$ to the vector $\sum_{\ks \in \cS} \alpha(\ks) \ks$.  Since every vector $\ket{\psi}$ has an expansion  $\ket{\psi} = \sum_{\vec{z} \in \Z_2^n} \psi_{\vec{z}} \ket{\vec{z}\,}$ in the computational basis, the map $e$ is surjective and the result follows by the rank-nullity theorem.
\end{proof}

We can immediately identify a basis for $\cL_n$.  Consider for each noncomputational stabiliser state $\ket{s} \in \cS_N$, its expansion in the computational basis $\ket{s} = \sum_{\vec{z} \in \Z_2^n} s_{\vec{z}} \ket{\vec{z}\,}$.  The corresponding linear dependency $\tilde s \in \C^{\cS}$ maps $\ket{s}$ to 1, $\ket{\vec{z}\,}$ to $-s_{\vec{z}}$ and all other stabiliser states to 0.  

\begin{lemma}The set of linear dependencies $\{\tilde s : \ks \in \cS_N\}$ is a basis for $\cL_n$.\end{lemma} \begin{proof}We only need to show that the $\tilde s$ are linearly independent. Consider a vanishing linear combination $\sum_{\ks \in \cS_N} c_{\ks} \tilde s = 0$ where 0 denotes the constant zero function.  Applying both sides of this equation, as functions in $\C^\cS$, to a noncomputational stabiliser state $\ket{t}$, we get that all $c_{\ket{t}} = 0$.\end{proof}  \noindent Working with this simple canonical basis is not particularly helpful.  This is because the maximum size of the support of each $\tilde s$ (the number of stabiliser states assigned a nonzero number by $\tilde s$) over all $n$-qubit $\ks$ grows exponentially in $n$.  The smallest-sized support of a nontrivial linear dependency is 3.  A natural question is whether it is possible to construct bases for $\cL_n$ consisting solely of such linearly dependent triples and, surprisingly, the answer is yes.

\section{A basis of triples}

Our strategy for generating a basis of the space of linear dependencies consisting solely of linearly dependent triples will be to decompose any given noncomputational stabiliser state $\ks$ as $\ks = 2^{-\frac{1}{2}}(\ket{t_1} + \ket{t_2})$, where $\ket{t_1}$ and $\ket{t_2}$ belong to the same stabiliser basis (and are thus orthogonal) and have support on precisely half as many computational basis states as $\ks$.  In order to do this, we require a characterisation of the size of the support of a stabiliser state.

\begin{thm}
Suppose $\ks$ is an $n$-qubit stabiliser state with Lagrangian subspace spanned by $\{(\vec{p}_j,\vec{q}_j) : j=1,\ldots,n\}$ and with eigenvalues captured by $\lambda_1,\ldots,\lambda_n \in \Z_2$.  Then $\ks$ has nonzero amplitudes on precisely $2^r$ computational basis states where $r$ is the rank of the $n \times n$ matrix over $\Z_2$ whose $j$-th row is $\vec{q}_j$.
%
\end{thm}

This is a consequence of the Dehaene-De Moor characterisation \cite[Theorem 5]{dehaene2003clifford} of stabiliser state amplitudes.  A simpler characterisation of the support of a stabiliser state, and proof of this theorem, is given in \cite{de2023fast}.

We are now prepared to give our decomposition of stabiliser states.  They yield examples of what Garcia et al. \cite{garcia2014geometry} term \emph{cofactors}: the postmeasurement states of a $Z$ measurement on an appropriately chosen qubit.

\begin{thm}[Splitting lemma]
Suppose $\ks$ is an $n$-qubit noncomputational stabiliser state with support on $2^r$ computational basis states.  There exist stabiliser states $\ket{t_1},\ket{t_2}$ specified by a common Lagrangian subspace with support on $2^{r-1}$ computational basis states such that $\ks = 2^{-\frac{1}{2}}(\ket{t_1} + \ket{t_2})$.
\end{thm}

\begin{proof}
For a given $\ket{s}$, there is a linearly independent set $\{(\vec{p}_j,\vec{q}_j) : j = 1,\ldots,n\}$ with pairwise vanishing symplectic products such that $\ks$ is a $(-1)^{\lambda_i}$-eigenvector of the Paulis $W(\vec{p}_j,\vec{q}_j)$.  We can assume without loss of generality that this set is chosen such that the the first $2n$ columns of the $n \times (2n + 1)$ \emph{check matrix} $$\begin{bmatrix}
\vec{q}_1 & \vec{p}_1 & \lambda_1\\
\vdots & \vdots & \vdots\\
\vec{q}_n &  \vec{p}_n & \lambda_n\\
\end{bmatrix}$$is in reduced row echelon form.  (Note that this matrix may also be called the \emph{tableau} of the state; however, this term may also refer to the same matrix with sign information omitted \cite{garcia2014geometry} or to the same matrix supplemented with destabiliser data \cite{aaronson2004improved}.)  This is because elementary row operations preserve linear independence of the rows, the property of pairwise vanishing symplectic inner products, and the Lagrangian subspace spanned by the rows.  Let $k$ be the position of the leading one in the first row.  Since $\ket{s}$ is not in the computational basis, $1 \leq k \leq n$.

Our aim is to replace $\vec{p}_1, \vec{q}_1$ with $\vec{P}, \vec{Q}$ to construct a new Lagrangian subspace such that the rank of the matrix whose rows are $\vec{Q},\vec{q}_2,\ldots,\vec{q}_n$ is $r-1$.  Assuming this has been established, we observe that $\ket{s}$ lies in the two-dimensional intersection of the $(-1)^{\lambda_j}$-eigenspaces of $W(\vec{p}_j,\vec{q}_j)$ for $j = 2, \ldots, n$.  This subspace is spanned by the two stabiliser vectors $\ket{t_1},\ket{t_2}$ defined by the new Lagrangian subspace and the sets of eigenvalues $$1, (-1)^{\lambda_2},\ldots,(-1)^{\lambda_n} \text{ and }{-1}, (-1)^{\lambda_2},\ldots,(-1)^{\lambda_n}$$ respectively.  Thus $\ks = \alpha \ket{t_1} + \beta{\ket{t_2}}$ for some $\alpha, \beta \in \C$.  As $\ket{t_1},\ket{t_2}$ each have support that is half the size of the support of $\ks$, the states $\ket{t_1},\ket{t_2}$ have disjoint support.  Since the amplitudes of $\ks$ and $\ket{t_1},\ket{t_2}$ all have absolute value $2^{-\frac{r}{2}}$ and $2^{-\frac{r-1}{2}}$ respectively, we see upon absorbing phases into the $\ket{t_j}$ that $\ks = 2^{-\frac{1}{2}}(\ket{t_1} + \ket{t_2})$ as desired.

We must now find $\vec{P}, \vec{Q}$ as described above.  Let \begin{align*}
C &= \{(\vec{p},\vec{q}) : [(\vec{p}, \vec{q}),(\vec{p}_j, \vec{q}_j)] = 0 \text{ for }j=2,\ldots,n\}\text{ and }\\
D &= \text{span}\{(\vec{p}_j, \vec{q}_j) : j=1,\ldots,n\}.
\end{align*}(Here, $\text{span}\{S\}$ denotes the linear subspace spanned by a subset $S$ of a vector space.)  Then an element of $\Z_2^{2n}$ completes $\{(\vec{p}_j, \vec{q}_j) : j = 2,\ldots,n\}$ to a basis for a new Lagrangian subspace if and only if it is a member of $C \setminus D$.  The set $C \setminus D$ is a coset of $D \subset \Z_2^{2n}$.

We claim that $(\vec{e}_k,\vec{0}) \in C \setminus D$.  Here, $\vec{e}_k$ denotes the $k$-th unit vector of $\Z_2^n$ and $(\vec{e}_k,\vec{0})$ corresponds to a Pauli with $Z$ in the $k$-th qubit: $\I \otimes \ldots \otimes Z \otimes \ldots \otimes \I$.  The vector $(\vec{e}_k,\vec{0})$ is in $C$ because $[(\vec{e}_k,\vec{0}),(\vec{p}_j, \vec{q}_j)]$ is the $k$-th component of $\vec{q}_j$ for $j=2,\ldots,n$; these components are zero since $k$ is the position of the leading one in the above matrix.   The vector $(\vec{e}_k,\vec{0})$ is not in $D$ since $[(\vec{e}_k,\vec{0}),(\vec{p}_1,\vec{q}_1)]$  is the $k$-th component of $\vec{q}_1$, i.e.\ takes the value 1, whereas every member of $D$ has vanishing symplectic product with $(\vec{p}_1, \vec{q}_1)$.  Thus, $\{(\vec{e}_k,\vec{0}),(\vec{p}_2, \vec{q}_2),\ldots,(\vec{p}_n, \vec{q}_n)\}$ generates a new Lagrangian subspace.  Choosing $(\vec{P}, \vec{Q}) = (\vec{e}_k,\vec{0})$ further ensures that the rank of the matrix whose rows are $\vec{0}, \vec{q}_2,\ldots,\vec{q}_n$ is $r-1$.\end{proof}

Applying this theorem to each noncomputational stabiliser state $\ks \in \cS_N$, we define the linear dependencies $D_{\ks}: \cS \to \C$ by $D_{\ks}(\ks)=1$, $\,D_{\ks}(\tau_j \ket{t_j}) = -2^{-\frac{1}{2}}\tau_j^{-1} $ for $j = 1,2$ and $\tau_j \in \C$ satisfying $\tau_j \ket{t_j} \in \cS$ (with $D_{\ks}$ assigning 0 to all other stabiliser states).  Forcing the $\ket{t_j}$ to have strictly smaller sized support as $\ks$ enables using these dependencies to construct a basis for all dependencies.

\begin{thm}
The set of linear dependencies $\{D_{\ks} : \ks \in \cS_N\} \subset \C^{\cS}$ is a basis for the space $\cL_n$ of all linear dependencies of $n$-qubit stabiliser states.
\end{thm}

\begin{proof}
Order the stabiliser vectors in $\cS$ ascendingly by the size of their support; the ordering among states with support of the same size can be chosen arbitrarily.  Consider the column vector representation of $D_{\ks}$ for a noncomputational stabiliser state $\ks$.  The column vector $D_{\ks}$ has nonzero entries only for $\ks$ and two other stabiliser vectors with support size strictly smaller than that of $\ks$.

Arrange these column vectors into a $|\cS| \times |\cS_N|$ matrix, using the same ordering of $\cS$ as before (restricted to the set of noncomputational stabiliser states).  The entries of this matrix are  $D_{\ket{s_2}}(\ket{s_1})$ for any $\ket{s_1} \in \cS$ and any $\ket{s_2} \in \cS_N$.  Since $D_{\ks}(\ks)=1$ for each noncomputational $\ks$ we see that each matrix entry along the lowest full diagonal takes the value 1 and, below this diagonal, each entry takes the value 0.  The matrix is thus full-rank and the $|\cS_N|$ column vectors representing the $D_{\ks} \in \cL_n$ are linearly independent.  Since the dimension of $\cL_n$ is also $|\cS_N|$, we conclude that the linear dependencies $D_{\ks}$ form a basis for $\cL_n$. \end{proof}

The construction of a basis for $\cL_n$ allows us to directly search the space of stabiliser decompositions of any state.

\section{Explicit computation of bases}

These bases for $\cL_n$ are useful for theoretically studying stabiliser decompositions.  However, for applications involving explicit optimisation, such as those outlined below, it is necessary to compute them.  

Here, we describe one way to generate the matrix $B$, which has as columns our basis of the space of linear dependencies, from a list of $n$-qubit Lagrangian subspaces.  This list is of exponentially smaller size than a dictionary of all stabiliser states which enables computations for higher numbers of qubits.     A Lagrangian subspace is specified by a basis, i.e.\ the data: $\{(\vec{p}_1,\vec{q}_1),\ldots,(\vec{p}_n,\vec{q}_n)\}$.

For convenience, we will make two assumptions about the basis data.  First, that the matrices
$$\begin{bmatrix}
   \quad \vec{q}_1 & & \vec{p}_1 \quad \\
   \quad \vdots & & \vdots \quad \\
   \quad \vec{q}_n & & \vec{p}_{n} \quad
\end{bmatrix},$$
are all in reduced row echelon form.  Second, that the list of Lagrangian subspaces has been sorted in increasing order of stabiliser state support size.  Recall that the support size of a stabiliser state is given by the rank of the matrix whose rows are $\vec q_i$. 

Next, for each Lagrangian subspace, we automatically get an ordering of its $2^n$ corresponding stabiliser states by listing the bit tuples $(\lambda_1, \ldots, \lambda_n)$ specifying eigenvalues in increasing order as binary integers. This defines an order on $\cS_N = \{\ket{s_1}, \ket{s_2}, \ldots\}$, and additionally an order on the whole of $\cS = \{\ket{\vec{0}}, \ldots, \ket{\vec{1}},\ket{s_1}, \ket{s_2}, \ldots\}$. We can keep track of this order by associating to each index the pairing of a basis and a bit tuple: $$\big(\{(\vec{p}_1,\vec{q}_1),\ldots,(\vec{p}_n,\vec{q}_n)\},\ (\lambda_1, \ldots, \lambda_n)\big).$$  This way, there is no need to calculate all of the amplitudes of any stabiliser state.  Recall that we fix the phase of each $\ket{s_j}$ by setting the amplitude corresponding to the smallest (as a binary integer) computational basis bitstring to be real and positive.

We will form a $|\cS| \times |\cS_N|$ matrix $B$ such that column $j$ corresponds to $D_{\ket{s_j}}$, i.e. the dependency $$\ket{s_j} - 2^{-\frac{1}{2}} \tau_{j,1}^{-1}\ket{t_{j,1}} - 2^{-\frac{1}{2}} \tau_{j,2}^{-1} \ket{t_{j,2}} = 0.$$  This column will have three nonzero entries.  In the index of the column corresponding to $\ket{s_j}$ will be a 1.  In the indices corresponding to the phase classes of $\ket{t_{j,1}}, \ket{t_{j,2}}$ will be numbers of the form $i^k 2^{-\frac{1}{2}}$.  Our goal will be to determine these numbers without generating any full states and position them appropriately in their column.

To work out the positions of the nonzero entries of column $j$, we take the Lagrangian subspace basis and bit tuple that correspond to $\ket{s_j}$ and modify the basis as per Theorem 2 to obtain the basis for the two split states $\ket{t_{j,1}}, \ket{t_{j,2}}$. We then row reduce this new basis, keeping track of how the bit tuples change, so that we obtain the canonical reduced row echelon form together with the correct pair of bit tuples.

Finally, it is possible to calculate the phases required, $\tau_{j,1}^{-1}$ and $\tau_{j,2}^{-1}$, without knowing the full stabiliser states. We determine the supports of $\ket{t_{j,1}}, \ket{t_{j,2}}$ using the method outlined below, then we identify the smallest bit string in each support. Denoting these bit strings by $\vec l_1, \vec l_2$, assume without loss of generality that $\vec l_1$ has the smaller integer value. Then we have $\tau_{j,1}^{-1} = 1$. To determine the phase of the other state, we look for a linear combination of the Lagrangian subspace basis for $\ket{s_j}$ that corresponds to a Pauli $P$ such that $P \ket{\vec l_1}$ is proportional to $\ket{\vec l_2}$. This can be done by first finding a linear combination $\vec q = \sum \alpha_j \vec q_j$ such that $\vec l_1 + \vec q = \vec l_2$. Then, we can use the coefficients $\alpha_j$ to find a product of the $W(\vec p_j, \vec q_j)$ that will give us the Pauli $P$ we want. The remaining phase is then given by $P \ket{\vec l_1} = \tau_{j,2}^{-1} \ket{\vec l_2}$.


To find the support of a stabiliser state, which is an affine subspace $\vec{c} + V = A \subset \Z_2^n$, given a Lagrangian subspace and bit tuple is to find a basis for the vector subspace $V \subset \Z_2^n$ and the vector $\vec{c} \in \Z_2^n$.  Here, we use the simple description of the support of a stabiliser state given in \cite{de2023fast}.  

If the basis for the Lagrangian subspace is in reduced row echelon form,
$$\begin{bmatrix}
   \quad \vec{q}_1 & & \vec{p}_1 \quad \\
   \quad \vdots & & \vdots \quad \\
   \quad \vec{q}_k & & \vec{p}_k \quad \\
   \quad \vec{0} & & \vec{r}_1 \quad \\
   \quad \vdots & & \vdots \quad \\
   \quad \vec{0} & & \vec{r}_{n-k}
\end{bmatrix},$$
then $V = \text{span}\{\vec{q}_1,\ldots,\vec{q}_k\}$. (If $k = n$ then the support is the whole of $\Z_2^n$ and we are done.) Furthermore, if the corresponding bit tuple is $(\lambda_1, \ldots, \lambda_k, \mu_1, \ldots, \mu_{n-k})$, then every vector $\vec{c} \in A$ must satisfy $\vec{r}_j \cdot \vec{c} = \mu_j$ for all $1\leq j\leq n-k$, where the dot product is taken over $\Z_2$. Any such $\vec{c}$ will satisfy $A = \vec{c} + V$.

\section{Applications}
Here, we demonstrate the utility of our results by computing the stabiliser extent of large states; see Table \ref{stabextenttable} below.  We then describe potential future applications to e.g.\ studying the stabiliser rank.

\paragraph{The stabiliser extent of large states}  The most extensive computations of the stabiliser extent \cite{beverland2020lower} have been limited to 5-qubit states.  This is because solving the optimisation problem $$
    \xi (|\psi\rangle) = \min{\lVert \vec x\,\rVert ^2_1}\ \text{ s.t. }|\psi\rangle = S_n \vec x
$$ requires a dictionary $S_n$ of all $n$-qubit stabiliser states and this is a set that grows like $2^{n^2/2}$.

We can recast this problem by minimising over the space of stabiliser decompositions:$$
    \xi (|\psi\rangle) = \min{\lVert \vec c + B\vec x\rVert ^2_1}.
$$  Here $\vec c$ is any choice of stabiliser decomposition of $\ket \psi$, e.g.\ the computational basis decomposition.  The matrix $B$ has as columns our basis of the space of linear dependencies, i.e.\ $\{D_{\ks} : \ks \in \cS_N\}$.  Although the change is subtle, our recasting is significantly more tractable than the standard approach for two reasons.  The first is that the matrix $B$ is incredibly sparse: every column has only 3 nonzero entries and, moreover, the matrix is lower triangular.  The second reason is that, as shown above, the matrix $B$ can be generated without first compiling a list of all stabiliser states.

We use this approach to compute the stabiliser extent of 6-qubit entangled states.  We consider the controlled-$Z$ gate magic state $\ket{C^5Z}$, which implement gates required for Grover search,
as the 5-qubit and smaller versions were studied in \cite{beverland2020lower}.  We also consider the Dicke states $\ket{D^6_k}$ which are defined as the uniform superposition of all 6-qubit computational basis states with Hamming weight $k$.  

For each of the 6-qubit states whose stabiliser extent we computed, the runtime of our program was just over 1000 seconds using only 32 CPUs.  We anticipate that it should be perfectly feasible to run our code for 7-qubit states without further optimisation and for larger states with further optimisation or theoretical simplifications.

The results are collected in Table \ref{stabextenttable}.

  
\renewcommand{\arraystretch}{2}
\begin{table}[h]
\centering
\begin{tabular}{|l|c|}
\hline
\textbf{State} & $\xi$  \\ \hline
$\ket{C^5Z}$ & $\frac{25}{16}$ \\ \hline
$\ket{D^6_1}$ & \multirow{2}{*}{$\frac{8}{3}$}  \\ \cline{1-1}
$\ket{D^6_5}$ &                   \\ \hline
$\ket{D^6_2}$ & \multirow{2}{*}{$\frac{12}{5}$} \\ \cline{1-1}
$\ket{D^6_4}$ &                   \\ \hline
$\ket{D^6_3}$ & $\frac{8}{5}$                  \\ \hline
\end{tabular}
\caption{The computed stabiliser extent of six-qubit states.}\label{stabextenttable}
\end{table}

\paragraph{Upper and lower bounds on the stabiliser rank}  We can take advantage of the newly revealed structure of stabiliser state geometry towards the design of algorithms to search for low-rank stabiliser decompositions.  Every stabiliser decomposition, including those of minimal stabiliser rank, can be derived from the standard computational basis decomposition by a sequence of additions of linearly dependent triples.  

This opens up a range of novel search strategies for low-rank stabiliser decompositions whose investigation is in progress.  It is possible to leverage known optimal stabiliser decompositions to study how they can be derived using discrete transformations on the standard decomposition.  We expect this approach to be more scalable than those based on random searches.

Finding lower-rank stabiliser decompositions leads directly to upper bounds on the stabiliser rank.  The practical impact of finding low-rank stabiliser decompositions will be to render more tractable the classical verification of prototype quantum computers via simulation.  This is because lowering the rank of stabiliser decompositions of magic states has the consequence of improving the exponential scaling of the stabiliser rank, and related, algorithms, thereby raising the number of qubits they can feasibly simulate \cite{qassim2021improved}.  Moreover, the insights gained by taking a structurally-informed approach to the construction of optimal stabiliser decompositions will illuminate the search for improved lower bounds on the stabiliser rank.  Once parsimonious patterns of transformations for generating optimal decompositions are identified, lower bounds will follow by establishing the moves' necessity.

\paragraph{Further applications}  Since stabiliser states are so ubiquitous in quantum information and given the obvious physical importance of linear combinations of states, i.e.\ superposition, deeper mathematical understanding of the structure of linear combinations of stabiliser states is certain to be widely relevant.  

Some potential contemporary domains of applicability beyond those discussed above include the resource theory of magic states \cite{howard2017application}, studies of the stabiliser polytope \cite{heinrich2019robustness}, and related classical simulation algorithms \cite{zurel2020hidden}.

\section{Acknowledgments}
ND acknowledges support from the Canada Research Chair program, NSERC Discovery Grant RGPIN-2022-03103, the NSERC-European Commission project FoQaCiA, and the Faculty of Science of Simon Fraser University.
SS acknowledges support from the Royal Society University Research Fellowship.
\section{Author declarations}

The authors have no conflicts to disclose.

\section{Data availability statement}

The code and data that support the findings of this study are publicly available at the following URL: \url{https://github.com/ndesilva/stabiliser-decomp-bases}.

\bibliographystyle{unsrt}
\bibliography{biblio}
\end{document}